\documentclass{article}
\usepackage{amsmath}
\usepackage{amssymb}
\usepackage{graphicx}

\begin{document}
\title{THE FINITE LAPLACE TRANSFORM FOR SOLVING A WEAKLY SINGULAR INTEGRAL EQUATION OCCURRING IN TRANSFER THEORY} 
\author{B. Rutily, L. Chevallier\\
Centre de Recherche Astronomique de Lyon\\(UMR 5574 du CNRS),\\
Observatoire de Lyon,\\ 9 avenue Charles Andr\'{e},\\ 69561 Saint-Genis-Laval Cedex, France\\
E-mail addresses: rutily@obs.univ-lyon1.fr (B. Rutily),\\ loic.chevallier@obs.univ-lyon1.fr (L. Chevallier)}
\date{Received 1 June 2004; Accepted 19 December 2004}
\maketitle

\abstract{We solve a weakly singular integral equation by Laplace transformation over a finite interval of $\mathbb{R}$. The 
equation is transformed into a Cauchy integral equation, whose resolution amounts to solving two Fredholm integral equations
of the second kind with regular kernels. This classical scheme is used to clarify the emergence of the auxiliary 
functions expressing the solution of the problem. There are four such functions, two of them being classical ones. This 
problem is encountered while studying the propagation of light in strongly scattering media such as stellar atmospheres.\\


\section{Introduction}

The integral form of the equation describing the radiative transfer of energy in a static, plane-parallel stellar 
atmosphere is \cite{ahuesetal2002a}
\begin{equation}
\label{eq1}
S(a,b,\tau)=S_0(a,b,\tau)+a\int_{0}^{b}K(\tau-\tau')S(a,b,\tau')d \tau',
\end{equation}
where $S$ is the source function of the radiation field and $S_0$ describes the radiation of the primary (internal or 
external) sources. These functions depend on the two parameters of the problem: the albedo $a\in]0,1[$, which characterizes the 
scattering properties of the stellar plasma, and the optical thickness $b>0$ of the atmosphere. They also depend on the optical 
depth $\tau\in[0,b]$, which is the space variable. Equation (\ref{eq1}) means that the radiation field at level $\tau$ is the
sum of the {\it{direct}} field from the primary sources, and the {\it{diffuse}} field having scattered at least once.\\
In the simplest scattering process conceivable - i.e., a monochromatic and isotropic one - the kernel of the integral equation (\ref{eq1})
 is the function
\begin{equation}
\label{eq2}
K(\tau):\,=\frac{1}{2}E_1(\vert \tau \vert)\quad(\tau\in \mathbb{R}^*),
\end{equation}
where $E_1$ is the first exponential integral function
\begin{equation}
\label{eq3}
E_1(\tau):\,=\int_{0}^{1}\exp(-\tau/x)\frac{dx}{x}\quad(\tau > 0).
\end{equation}
Since $E_1(\vert \tau \vert) \sim - \ln(\vert \tau \vert)$ when $\vert \tau \vert \to 0^+$, the kernel of the integral equation 
(\ref{eq1}) is weakly singular on its diagonal. The free term $S_0$ includes the thermal emission of the stellar
 plasma - of the form $(1-a)B^*(\tau)$, where $B^*$ is a known function - and the contribution $aJ_0^{ext}(b,\tau)$ of the external sources
 via the boundary conditions: see the introduction of \cite{ahuesetal2002a}. Hence $S_0(a,b,\tau)=(1-a)B^*(\tau)+aJ_0^{ext}(b,\tau)$. 
In a homogeneous and isothermal atmosphere assumed to be in local thermodynamic equilibrium, the function $B^*$ coincides with the 
Planck function $B(T)$ at the (constant) temperature $T$. Moreover, $J_0^{ext}=0$ in the absence of external sources and thus 
\begin{equation}
\label{eq4} 
S_0(a,b,\tau)=(1-a)B(T),
\end{equation}
which shows that $S_0$ is independent of $\tau$ in this model. The solution $S$ to the problem (\ref{eq1}) is then
\begin{equation}
\label{eq5}
S(a,b,\tau)=(1-a)B(T)Q(a,b,\tau),
\end{equation}
where $Q$ solves the following integral equation:
\begin{equation}
\label{eq6}
Q(a,b,\tau)=1+a\int_{0}^{b}K(\tau-\tau')Q(a,b,\tau')d \tau'.
\end{equation}
It is proved in \cite{ahuesetal2002b} that the space $C^0([0,b])$ of the continuous functions from $[0,b]$ to $\mathbb{R}$
 is invariant under the operator
\begin{equation}
\label{eq7}
\Lambda\;:\;f\;\;\rightarrow\;\;\Lambda f(\tau):\,= \int_{0}^{b}K(\tau-\tau')f(\tau')d\tau',
\end{equation}
with norm
\begin{equation}
\label{eq8}
\|\Lambda\|_{\infty}= \int_{0}^{b/2}E_1(\tau)d\tau=1-E_2(b/2).
\end{equation}
Here, $E_2(\tau):=\int_{0}^{1}\exp(-\tau/x)dx$ is the exponential integral function of order 2. Equation (\ref{eq6}), 
which can be written in the form
\begin{equation}
\label{eq9}
Q=1+a \Lambda Q,
\end{equation}
has therefore a unique solution in $C^0([0,b])$ provided that $a<1$ or $b<+\infty$.\\
This problem is a basic one in stellar atmospheres theory, and more generally in transport theory. It describes in the simplest way 
the multiple scattering of some type of particles (here photons) on scattering centers distributed uniformly in a slab of finite
 thickness, a very simple 1D-configuration. Its applications in astrophysics and neutronics - among other fields - are presented in 
\cite{rutily2002}. It is important to solve this problem very accurately, thus providing a benchmark to validate 
the numerical solutions of integral equations of the form (\ref{eq1}).\\
Physicists and astrophysicists have developed many methods for solving integral equations of the form (\ref{eq1}) 
with a convolution kernel defined by (\ref{eq2}) \cite{rutily-bergeat1994}. The main steps for solving the prototype equation 
(\ref{eq6}) are summarized in a recent article \cite{chevallier-rutily2004}, which contains accurate tables of the 
function $(1-a)Q$ for different values of the parameters $a$ and $b$. While reading this paper, one is struck by the
complexity of the ``classical'' solution to Eq. (\ref{eq6}). It requires the introduction of many intricate auxiliary 
functions introduced in the literature over more than thirty years. The reader quickly loses the thread of the solution, which 
reduces his chances of exploiting it for solving problems of the more general form (\ref{eq1}).\\ 
The aim of the present article is to get around this difficulty by solving Eq. (\ref{eq6}) straightforwardly, introducing as few 
auxiliary functions as possible to express its solution. These functions are briefly studied in Appendix A. The method is
based on the finite Laplace transform, which reduces the problem (\ref{eq6}) to solving a Cauchy integral equation over 
$[-1, +1]$, which in turn can be transformed into two Fredholm integral equations over [0, 1]. This approach has been 
developed in transport theory after the publication in 1953 of the first English translation of Muskhelishvili's 
monograph {\it Singular integral equations} \cite{muskhelishvili1992}: see, e.g., \cite{busbridge1955}, \cite{mullikin1964} and
\cite{carlstedt-mullikin1966}. It can be considered as an extension of the Wiener-Hopf method \cite{busbridge1960} for solving 
integral equations of the form (\ref{eq1}) with $b<\infty$. Both methods are characterized by an intensive use of the theory 
of (sectionally) analytic functions, which allows solution of Eq. (\ref{eq1}) in a concise manner. This is obvious
 when comparing the solution we derive here to the classical solution of the particular problem (\ref{eq6}), which does not use 
any calculation in the complex plane. The former method clarifies the origin and the role of the auxiliary functions
 expressing the solution to a problem of the form (\ref{eq1}). Since these functions are independent of the source term 
$S_0$, they are ``universal'' for a given scattering kernel.\\
The remainder of this article is organized as follows: in Sec. 2, the finite Laplace transform of the $Q$-function is calculated 
on the basis of some recent developments on Cauchy integral equations \cite{rutily-bergeat2002}-\cite{rutily-chevallier-bergeat2004}. 
Then the Laplace transform is inverted and the solution achieved in \cite{chevallier-rutily2004} is concisely retrieved with 
the help of the theorem of residues (Sec. 3). It involves two functions $F_+$ and $F_-$ with remarkable properties, as shown in 
Appendix B. The difficulties arising from the numerical evaluation of the latter functions are investigated in \cite{chevallier-rutily2004}.  

\section{The calculation of the finite Laplace transform of the $Q$-function}

Supposing $0<b<\infty$, we plan to solve Eq. (\ref{eq9}) by Laplace transform (LT) over $[0,b]$. This operator is defined on $C^0([0,b])$ by
\begin{equation}
\label{eq10}
\mathfrak{L}f(z):=\int_{0}^{b}f(\tau)\exp(-\tau z)d\tau\quad(z\in \mathbb{C}).
\end{equation}
Since $b<\infty$, the finite LT of a continuous function is defined and analytic in the whole complex plane. The inversion formula
\begin{equation}
\label{eq11}
f(\tau)=\frac{1}{2i\pi} -\hspace{-1.05em}\int_{c-i\infty}^{c+i\infty} \mathfrak{L}f(z)\exp(\tau z)dz
\end{equation}
is valid at any $\tau\in\,]0,b[$, with no restriction on $c\in\mathbb{R}$. The symbol $ -\hspace{-0.91em}\int$ on the right-hand side of Eq. (\ref{eq11}) 
means that the integral is a Cauchy principal value at infinity, i.e., $\displaystyle -\hspace{-1.05em}\int_{c-i\infty}^{c+i\infty} = \lim_{X\to+\infty} \int\limits_{c-iX}^{c+iX}$.\\
With the intention of taking the LT of both members of Eq. (\ref{eq9}), we note that the LT of the function $K$, as defined by
 Eqs. (\ref{eq2})-(\ref{eq3}), exists on $\mathbb{C}\setminus\lbrace\pm1\rbrace$ and is
\begin{equation}
\label{eq12}
\mathfrak{L}K(z)=w(1/z)-\frac{1}{2}\int_{0}^{1}\frac{du}{1-z u}-\exp(-bz)\frac{1}{2}\int_{0}^{1}\exp(-b/u)\frac{du}{1+zu},
\end{equation}
where $w:\,\mathbb{C}\setminus\lbrace\pm1\rbrace \;\rightarrow \;\mathbb{C}$ denotes the function
\begin{equation}
\label{eq13}
w(z):\,=\frac{z}{2}\int_{-1}^{+1}\frac{du}{u+z}.
\end{equation}
The three integrals on the right-hand side of Eqs. (\ref{eq12})-(\ref{eq13}) are Cauchy principal values over $]1,+\infty[$, 
$]-\infty,-1[$ and $]-1, +1[$, respectively. From the definition (\ref{eq7}) of the $\Lambda$-operator, we infer that the 
finite LT of $\Lambda f$ is
\begin{multline}
\label{eq14}
\mathfrak{L}({\Lambda f})(z)=w(1/z)\mathfrak{L}f(z)-\frac{1}{2}\int_{0}^{1}\mathfrak{L}f(1/u)\frac{du}{1-zu}\\-\exp(-bz)\frac{1}{2}\int_{0}^{1}\mathfrak{L}f(-1/u)\exp(-b/u)\frac{du}{1+zu}.
\end{multline}    
In addition, the finite LT of the unit function is $z\to(1/z)[1-\exp(-bz)]$. Taking the LT of both members of Eq. (\ref{eq9}) and 
changing $z$ into $1/z$, we obtain the following integral equation for $\mathfrak{L}Q$:
\begin{multline}
\label{eq15}
T(a,z)\mathfrak{L}Q(a,b,1/z)-\frac{a}{2}z\int_{0}^{1}\mathfrak{L}Q(a,b,1/u)\frac{du}{u-z}\\+\exp(-b/z)\frac{a}{2}z\int_{0}^{1}\mathfrak{L}Q(a,b,-1/u)\exp(-b/u)\frac{du}{u+z}=c_0(z),
\end{multline}
where
\begin{equation}
\label{eq16}
T(a,z):\,=1-aw(z)=1-\frac{a}{2}z \int_{-1}^{+1}\frac{du}{u+z}
\end{equation}
and
\begin{equation}
\label{eq17}
c_0(z):\,=z[1-\exp(-b/z)].
\end{equation}
The function $T$ is our first basic auxiliary function. Its main properties are summarized in Appendix A. The function $c_0$ 
satisfies the three following conditions: ({\it{i}}) it is defined and analytic in $\mathbb{C}^*$, 
({\it{ii}}) it is bounded at infinity (with limit equal to $b$), and ({\it{iii}}) it satisfies, in a neighborhood of 0, the 
conditions
\begin{equation}
\label{eq18}
\lim_{\scriptstyle z\to 0 \atop \scriptstyle \mathrm{Re}(z)>0}c_0(z) < \infty\;\;,\;\lim_{\scriptstyle z\to 0 \atop \scriptstyle \mathrm{Re}(z)>0}[c_0(-z)\exp(-b/z)] < \infty.
\end{equation}
Replacing $u$ by $-u$ in the second integral of Eq. (\ref{eq15}), we obtain a Cauchy integral equation on $[-1, +1]$ for the function 
$z \to \mathfrak{L}Q(a,b,1/z)$. Since the latter is not H\"older-continuous at 0, it seems inappropriate to undertake 
the resolution of this equation. A better approach consists in observing that the functions $z \to \mathfrak{L}Q(a,b,1/z)$ and
$z \to \mathfrak{L}Q(a,b,-1/z)$ $\times\exp(-b/z)$ are solutions of two coupled Cauchy integral equations on $[0, 1]$, which can be uncoupled 
by adding and subtracting them. The resulting Cauchy integral equations are then reduced to two Fredholm integral equations of the second kind
on $[0, 1]$ with regular kernels. This general approach was first introduced in transfer theory by Busbridge \cite{busbridge1955} and 
developed by Mullikin et al. \cite{mullikin1964,carlstedt-mullikin1966} and Rutily et al. \cite{rutily-chevallier-bergeat2004}.
The last mentioned reference gives a synthesis that the reader may consult for details. It contains the proof of the 
following result, which we omit here: the unique solution, analytic in $\mathbb{C}^*$, to an integral equation of the form (\ref{eq15}) with free 
term satisfying the conditions ({\it{i}})-({\it{iii}}), can be written in the form    
\begin{equation}
\label{eq19}
\mathfrak{L}Q(a,b,1/z)=\frac{1}{2}[u_-(a,b,z)\eta_{0,+}(a,b,z)+u_+(a,b,z)\eta_{0,-}(a,b,z)], 
\end{equation}
where, for any $z\in \mathbb{C}\setminus i\mathbb{R}$,
\begin{multline}
\label{eq20}
u_{\pm}(a,b,z):\,= \mathrm{Y}[\Re(z)]H(a,z)\zeta_{\pm}(a,b,-z)\\ \mp \mathrm{Y}[-\Re(z)]H(a,-z)\zeta_{\pm}(a,b,z)\exp(-b/z),
\end{multline} 
and
\begin{multline}
\label{eq21}
\eta_{0,\pm}(a,b,z):\,=\frac{1}{2}c_{0,\mp}(a,b,+0)+\frac{z}{2i\pi}\int_{-i\infty}^{+i\infty}H(a,1/z')\\ \times (\zeta_{\pm})^-(a,b,-1/z')c_0(-1/z')\frac{dz'}{1+zz'}.
\end{multline}
Equation ({\ref{eq20}) involves the Heaviside function Y, which vanishes over $\mathbb{R}_{-}^{*}$ and is equal to unity over 
$\mathbb{R}_{+}^{*}$. On the right-hand side of Eq. (\ref{eq21}), we have introduced
\begin{equation}
\label{eq22}
c_{0,\mp}(a,b,+0):=\lim_{\scriptstyle z\to 0 \atop \scriptstyle \mathrm{Re}(z)>0}[c_0(z)\pm c_0(-z)\exp(-b/z)].
\end{equation}
Since this limit vanishes for the particular $c_0(z)$ given by (\ref{eq17}), Eq. (\ref{eq21}) can be rewritten as
\begin{multline}
\label{eq23}
\eta_{0,\pm}(a,b,z)=-\frac{1}{2i\pi}\int_{-i\infty}^{+i\infty}H(a,1/z')(\zeta_{\pm})^-(a,b,-1/z')\\ \times [1-\exp(bz')]\frac{dz'}{z'(z'+1/z)}.
\end{multline}
The functions $H(a,z)$, $\zeta_+(a,b,z)$ and $\zeta_-(a,b,z)$ appearing in Eqs. (\ref{eq20}) and (\ref{eq23}) are the 
other three auxiliary functions of our approach. They are defined in Appendix A, which synthesizes their main properties. The 
$H$-function is a classical function of radiation transfer theory, recently reviewed in \cite{rutily-bergeat-chevallier2004}. 
Most results concerning the functions $\zeta_{\pm}$ are summarized in \cite{rutily-chevallier-bergeat2004}. They are sectionally analytic in the complex plane cut along 
the imaginary axis, with limits $(\zeta_{\pm})^+$ and $(\zeta_{\pm})^-$ on the right- and left-hand sides of this axis respectively. 
These limits satisfy the relation
\begin{equation}
\label{eq24}
H(a,z_0)(\zeta_{\pm})^-(a,b,-z_0)=\mp H(a,-z_0)(\zeta_{\pm})^-(a,b,z_0)\exp(-b/z_0),
\end{equation}
for any $z_0\in i\mathbb{R}^*$. This can be seen by putting $z \to z_0$ on the right, then on the left, into Eq. (\ref{eq20}),
taking into account the continuity of the functions $u_{\pm}$ on $i\mathbb{R}$ \cite{rutily-chevallier-bergeat2004}. The limits
of the functions $\zeta_{\pm}$ on both sides of the imaginary axis also satisfy the relation
\begin{multline}
\label{eq25}
H(a,z_0)(\zeta_{\pm})^+(a,b,-z_0)=H(a,z_0)(\zeta_{\pm})^-(a,b,-z_0)\\ \pm H(a,-z_0)(\zeta_{\pm})^+(a,b,z_0) \exp(-b/z_0),
\end{multline}
which follows from (\ref{eqA11}) and Plemelj's formulae \cite{muskhelishvili1992}.\\
In the expression (\ref{eq23}) of the functions $\eta_{0,\pm}$, replace the term containing $\exp(bz')$ by its expression 
given by (\ref{eq24}) (with $z_0=-1/z'$). One obtains
\begin{multline}
\label{eq26}
\eta_{0,\pm}(a,b,z)=-\frac{1}{2i\pi}\int_{-i\infty}^{+i\infty}H(a,1/z')(\zeta_{\pm})^-(a,b,-1/z')\\ \times [\frac{1}{z'+1/z}\pm\frac{1}{z'-1/z}]\frac{dz'}{z'}.
\end{multline}
\begin{figure}[htb]
\centering
\includegraphics{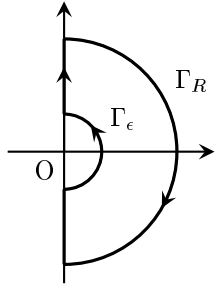}
\label{fig1}
\caption{Contour for the calculation of $\eta_{0,\pm}(a,b,z)$.}
\end{figure}
This integral may be calculated using the contour of Fig. 1. The integral along the half-circle $\Gamma_{\epsilon}$ tends to 
$-(1/2)H(a,\infty)\zeta_{\pm}(a,b,-\infty)(z\mp z)$ as $\epsilon \to 0$, where $H(a,\infty)$ is the limit of $H(a,z)$
when $z \to \infty$ in any part of the complex plane. This limit is given by Eq. (\ref{eqA9}). $\zeta_{\pm}(a,b,-\infty)$ denotes
the limit of $\zeta_{\pm}(a,b,z)$ when $z \to \infty$ in the left complex half-plane. In Eq. (\ref{eqA18}), this limit is expressed 
in terms of $\zeta_{\pm}(a,b,+\infty)$. The integral along $\Gamma_R$ tends to 0 when $R \to \infty$. From the theorem of residues,
we have for any $z\in \mathbb{C}\setminus i\mathbb{R}$
\begin{multline}
\label{eq27}
\eta_{0,\pm}(a,b,z)=\frac{z}{\sqrt{1-a}}(1\mp 1)(\zeta_{\pm})(a,b,+\infty)\\
-z \left\lbrace \mathrm{Y}[-\Re(z)]H(a,-z)\zeta_{\pm}(a,b,z)\mp \mathrm{Y}[\Re(z)]H(a,z)\zeta_{\pm}(a,b,-z) \hspace{-0.8em}\phantom{\frac{a}{a}}\right\rbrace .
\end{multline}
The end of the calculation of $\mathfrak{L}Q$ follows from Eqs. (\ref{eq19}), (\ref{eq20}) and (\ref{eq27}). Changing $z$ 
into $1/z$, we get 
\begin{equation}
\label{eq28}
\mathfrak{L}Q(a,b,z)=\frac{1}{\sqrt{1-a}}\frac{1}{z}\zeta_-(a,b,+\infty)u_+(a,b,1/z),
\end{equation}
a result valid in principle on $\mathbb{C}\setminus i\mathbb{R}$. In fact, it can easily be seen that the limits of 
this function on both sides of the imaginary axis coincide. This makes it possible to extend the relation (\ref{eq28})
 by continuity on $i\mathbb{R}$, including 0. Such an extension will not be used subsequently.

\section{Inverting the Laplace transform}
The inversion formula (\ref{eq11}) reads here, for any $\tau\in]0,b[$
\begin{equation}
\label{eq29}
Q(a,b,\tau)=\frac{1}{2 i\pi}\int_{c-i\infty}^{c+i\infty}\mathfrak{L}Q(a,b,z)\exp(\tau z)dz
\end{equation}
($c\not=0$). Substituting (\ref{eq28}) in this expression and replacing $u_+(a,b,1/z)$ by its definition (\ref{eq20}), we
obtain
\begin{multline}
\label{eq30}
Q(a,b,\tau)=\frac{1}{\sqrt{1-a}}\zeta_-(a,b,+\infty)\\\times \frac{1}{2i\pi}\int_{c'-i\infty}^{c'+i\infty}H(a,1/z)\zeta_+(a,b,-1/z)\exp(\tau z)\frac{dz}{z}, 
\end{multline}
where $c'=\vert c \vert>0$. The integral on the right-hand side can be calculated with the help of the residue 
theorem applied to a contour in the left half-plane, owing to the presence of the exponential. A difficulty arises
because $c'>0$ and the imaginary axis should not be crossed since the functions $\zeta_{\pm}$ are discontinuous on 
this axis. Nevertheless, one can derive the identity
\begin{multline}
\label{eq31}
\frac{1}{2i\pi}\int_{c'-i\infty}^{c'+i\infty}H(a,1/z)\zeta_+(a,b,-1/z)\exp(\tau z)\frac{dz}{z}=
\\\frac{1}{2i\pi}\,-\hspace{-1.1em}\int_{-i\infty}^{+i\infty} H(a,1/z)(\zeta_+)^-(a,b,-1/z)\exp(\tau z)\frac{dz}{z} 
\end{multline}
from the contour of Fig. 2(a), since the integral on the half-circle $\Gamma_{\epsilon}$ tends to 0 when $\epsilon \to 0$, due 
to Eq. (\ref{eqA18}). The last integral is a Cauchy principal value at 0. Then the limits of the functions $\zeta_{\pm}$ 
on the left-hand side of the imaginary axis are expressed in terms of their limits on the right-hand side 
with the help of (\ref{eq25}), which shows that
\begin{multline}
\label{eq32}
\frac{1}{2i\pi}\,-\hspace{-1.075em}\int_{-i\infty}^{+i\infty} H(a,1/z)(\zeta_+)^-(a,b,-1/z)\exp(\tau z)\frac{dz}{z}=
\\\frac{1}{2i\pi}\,-\hspace{-1.075em}\int_{-i\infty}^{+i\infty} H(a,1/z)(\zeta_+)^+(a,b,-1/z)[\exp(\tau z)+\exp((b-\tau)z)]\frac{dz}{z}. 
\end{multline}
\begin{figure}[htb]
\centering
\includegraphics{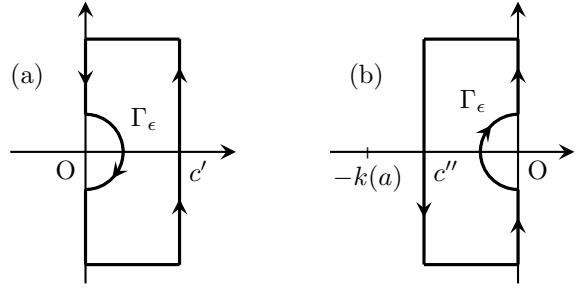}
\label{fig2}
\caption{Contours (a) and (b) for the proof of Eqs. (\ref{eq31}) and (\ref{eq33}).}
\end{figure}
The last integral is calculated by displacing the line of integration on the left side of the imaginary axis, between the origin and the point of abscissa $-k(a)$ where the function 
$z\to H(a,1/z)$ diverges: see the comments after Eq. (\ref{eqA8}) in Appendix A, which contains the definition of $k(a)$. 
Applying the residue theorem to the contour of Fig. 2(b) and observing that the integral on $\Gamma_{\epsilon}$ tends to 
$-H(a,\infty)\zeta_+(a,b,+\infty)$ when $\epsilon \to 0$, one obtains
\begin{align}
\label{eq33}
\frac{1}{2i\pi}& -\hspace{-1.1em}\int_{-i\infty}^{+i\infty} H(a,1/z)(\zeta_+)^+(a,b,-1/z)[\exp(\tau z)+\exp((b-\tau)z)]\frac{dz}{z} \nonumber\\
&=\frac{1}{\sqrt{1-a}}\zeta_+(a,b,+\infty) \\
&+\frac{1}{2i\pi}\int_{c''-i\infty}^{c''+i\infty}H(a,1/z)\zeta_+(a,b,-1/z)[\exp(\tau z)+\exp((b-\tau)z)]\frac{dz}{z},\nonumber
\end{align}
where $c''\in\,]-k(a),0[$. For $\tau\in\,]0,b[$, the last integral coincides with the function $-F_+(a,b,\tau)$ introduced in 
the Appendix B, Eq. (\ref{eqB1}).\\
According to Eqs. (\ref{eq30})-(\ref{eq33}), we finally have
\begin{equation}
\label{eq34}
Q(a,b,\tau)=\frac{1}{\sqrt{1-a}}\,\zeta_-(a,b,+\infty)[\frac{1}{\sqrt{1-a}}\,\zeta_+(a,b,+\infty)-F_+(a,b,\tau)],
\end{equation}  
or, using Eq. (\ref{eqB3})
\begin{equation}
\label{eq35}
Q(a,b,\tau)=[1+F_-(a,b,0)][1+F_+(a,b,0)-F_+(a,b,\tau)].
\end{equation}
This expression is valid {\it a priori} over $]0,b[$, but it can be extended by continuity over $[0,b]$, since the functions $F_{\pm}$
 are continuous on the right at 0 and on the left at $b$. One has in particular
\begin{equation}
\label{eq36}
Q(a,b,0)=Q(a,b,b)=1+F_-(a,b,0).
\end{equation} 
These results are similar to those of \cite{chevallier-rutily2004}, where details concerning the numerical evaluation of
the function $Q$ can be found. Ten-figure tables of the function $(1-a)Q$ are given there. The functions $F_{\pm}$ 
are computable once their definition (\ref{eqB1}) has been transformed by the method of residues, as outlined in Appendix B.
\section{Conclusion}
Using the finite Laplace transform, we derived in a concise way the expression (\ref{eq35}) of the solution to 
Eq. (\ref{eq6}), which is appropriate for numerical evaluation \cite{chevallier-rutily2004}. 
Our main objective was to come directly to this expression, with emphasis on the manner in which the four auxiliary 
functions of the problem were generated. They are ({\it{i}}) the function $T=T(a,z)$, which characterizes the solution 
to Eq. (\ref{eq6}) in an infinite medium (the range $[0,b]$ is replaced by $\mathbb{R}$), ({\it{ii}}) the function $H=H(a,z)$
 which expresses its solution in a half-space ($b=\infty$), and ({\it{iii}}) the functions $\zeta_{\pm}=\zeta_{\pm}(a,b,z)$ 
that complete the previous ones in the finite case ($b<\infty$). The main properties of these functions are summarized in 
Appendix A. We note that the $T$-function can be expressed in terms of elementary transcendental functions [Eq. (\ref{eqA3})],
 the $H$-function is defined {\it{explicitly}} by an integral on $[0,1]$ [Eq. (\ref{eqA10})], and the functions $\zeta_{\pm}$ 
are defined {\it{implicitly}} as the solutions to Fredholm integral equations of the second kind [Eq. (\ref{eqA13})]. It 
seems that the problem (\ref{eq6}) has no exact closed-form solution for $b<\infty$.\\
The approach of this article is appropriate for solving integral equations of the form (\ref{eq1}), with a kernel 
defined by (\ref{eq2}) and any free term. It also applies to convolution kernels defined by functions more general than 
(\ref{eq2}), for example of the form
\begin{equation}
\label{eq37}
K(\tau):=\int_{I}\Psi(x)\exp(-\vert \tau \vert /x)dx,
\end{equation} 
where $I$ is an interval of $\mathbb{R}$ and $\Psi$ a function ensuring the existence of the integral. This class of kernels 
models scattering processes more complex than the one considered in this article, which corresponds to $I=\,]0,1]$ and $\Psi(x)=(1/2)(1/x)$.

\appendix
\renewcommand{\theequation}{\Alph{section}\arabic{equation}}
\setcounter{equation}{0}
\setcounter{section}{1}

\section*{Appendix A. The auxiliary functions associated to the problem (\ref{eq6})}

These functions are $T=T(a,z)$, $H=H(a,z)$, and $\zeta_{\pm}=\zeta_{\pm}(a,b,z)$, where $z\in\mathbb{C}$.\\
\subsection{The $T$-function}
This function is defined by
\begin{equation}
\label{eqA1}
T(a,z):=1-{\frac{a}{2}}z\int_{-1}^{+1}{\frac{du}{u+z}}\qquad(z\neq\pm1).
\end{equation}
It describes the multiple scattering of photons in an unbounded medium for the adopted scattering law. It is defined in the whole 
complex plane $\mathbb{C}$, except at $\pm 1$, provided that the integral is calculated 
in the sense of the Cauchy principal value when $z \in \,]-1,+1[$. It is thus sectionally analytic in $\mathbb{C}\setminus[-1,+1]$,
its limits on both sides of the segment $]-1, +1[$ being given from Plemelj formulae \cite{muskhelishvili1992} by 
\begin{equation}
\label{eqA2}
T^{\pm}(a,u)=T(a,u)\pm i\pi(a/2)u\quad(-1<u<+1).
\end{equation}
Here, $T^+(a,u)$ [resp. $T^-(a,u)$] denotes the limit of $T(a,z)$ when $z$ tends to $u\in\,]-1,+1[$ from above (resp. below) the real axis, 
and
\begin{equation}
\label{eqA3}
T(a,u)=1-\frac{a}{2}u \ln(\frac{1+u}{1-u})\quad(-1<u<+1).
\end{equation}
The roots of the characteristic equation $T(a,z)=0$ are important for solving integral equations of the form (\ref{eq1}) with 
kernel defined by (\ref{eq2}). When $0<a<1$, this equation has four non-zero roots in $\mathbb{C}$, namely two pairs of 
opposite real numbers since the $T$-function is even \cite{busbridge1960}. There is a unique root strictly greater than unity, denoted 
by $1/k(a)$ ($0<k(a)<1$), which is calculated by solving the transcendental equation
\begin{equation}
\label{eqA4}
T(a,1/k(a)) = 1 - {\frac{a}{2}}{\frac{1}{k(a)}} \ln \left[\frac{1+k(a)}{1-k(a)}\right] = 0 .
\end{equation}
Its unique solution $k(a)$ in ]0,1[ is given by \cite{carlstedt-mullikin1966}
\begin{equation}
\label{eqA5}
k(a) = \sqrt{1-a} \, \exp \left[\int_{0}^{1} \theta (a,u)\frac{du}{u}\right],
\end{equation}
where
\begin{equation}
\label{eqA6}
\theta (a, u) := \frac{1}{\pi} \arg[T^+(a,u)] \quad (0 \leq u < 1).
\end{equation}
Here, $\arg(z)$ is the principal value of the argument of $z\in\mathbb{C}$. It can be computed using the atan2-function, since
$\arg(x+iy)=\mathrm{atan2}(y,x)$.
\subsection{The $H$-function}
This function is the unique solution, analytic in the right complex half-plane, of the integral equation
\begin{equation}
\label{eqA7}
T(a,z)H(a,z) = 1 + \frac{a}{2}z\int_{0}^{1}H(a,u)\frac{du}{u-z} \quad(z\not=\pm1).
\end{equation}
This is a Cauchy integral equation on $[0,1]$, which defines the extension of $H$ outside this interval. The function 
$H$ satisfies the Wiener-Hopf factorization relation
\begin{equation}
\label{eqA8}
T(a,z)H(a,z)H(a, -z)=1 \quad(z\not=\pm1),
\end{equation} 
first proved regardless of (\ref{eqA7}) (see, e.g., \cite{busbridge1960}), then as a result of (\ref{eqA7}) 
\cite{rutily-bergeat-chevallier2004}. We note that it implies the divergence of the $H$-function at the solutions of the
equation $T(a,z) = 0$ located in the left complex half-plane, especially at $z=-1/k(a)$. The relations (\ref{eqA7})-(\ref{eqA8}) 
also mean that the $H$-function is bounded at infinity, with limit
\begin{equation}
\label{eqA9}
H(a,\infty)=\frac{1}{\sqrt{1-a}}
\end{equation} 
whatever the region of the complex plane where $z$ goes to infinity. This results from the fact that $T(a,\infty)=1-a$.\\
Many analytical expressions of the function $H$ have been derived since the introduction of this function in the thirties. 
The following expression is valid in the right complex half-plane \cite{carlstedt-mullikin1966}
\begin{equation}
\label{eqA10}
H (a,z) = \frac{k(a)}{\sqrt{1-a}}\frac{1+z}{1+k(a)\,z}\exp \left[-z\int_{0}^{1}\theta(a,u)\frac{du}{u+z}\right]\; (\Re(z)\geq 0),
\end{equation}
where $\theta$ is the function defined by (\ref{eqA6}). We used it to compute $H(a,z)$ over $[0,1]$ and at $1/k(a)$,
a necessary step for computing the functions $F_{\pm}$ appearing in our final result (\ref{eq35}): see the Appendix B.

\subsection{The functions $\zeta_{\pm}$}
These functions are sectionally analytic in $\mathbb{C}\setminus i\mathbb{R}$, with limits $(\zeta_{\pm})^+$ and $(\zeta_{\pm})^-$ 
on the right- and left-hand sides of the imaginary axis, respectively. They are defined by the relations
\begin{eqnarray}
\label{eqA11}
\lefteqn{\zeta_{\pm}(a,b,z)=1\pm \frac{z}{2i\pi}\int_{-i\infty}^{+i\infty}\frac{H(a,-1/z')}{H(a,1/z')}}\nonumber\\
&&\qquad\qquad\qquad\qquad\quad\times\exp(-bz')(\zeta_{\pm})^+(a,b,1/z')\frac{dz'}{1\!+\!zz'},
\end{eqnarray}
which can be transformed into Fredholm integral equations of the second kind on $[0,1]$ using the residue theorem 
\cite{rutily-chevallier-bergeat2004}. We suppose here the existence and uniqueness of the solution to Eq. (\ref{eqA11}), 
and refer to \cite{rutily-chevallier-bergeat2004} for the proof of the relation 
\begin{equation}
\label{eqA12}
\frac{1}{2}[\zeta_+\!(a,b,z)\zeta_-\!(a,b,-z)+\zeta_-\!(a,b,z)\zeta_+\!(a,b,-z)]=1, 
\end{equation}
which is valid over $\mathbb{C}\setminus i\mathbb{R}$.\\
The transformation of Eq. (\ref{eqA11}) into a couple of Fredholm integral equations on $[0,1]$ is described in the section 
V of \cite{rutily-chevallier-bergeat2004}. We reproduce the relation (71) of this article, valid for any $z$ in $\mathbb{C}
\setminus\lbrace\,]-1,0[\,\cup\,\lbrace-1/k(a)\rbrace\rbrace$
\begin{eqnarray}
\label{eqA13}
\zeta_{\pm}(a,b,z)&=&1\pm M_{\pm}(a,b)\frac{2kz}{1+kz}\nonumber\\&\mp&{\rm Y}[-\Re(z)]\frac{1}{T(a,z)}\frac{\exp(b/z)}{H^{2}(a,-z)}\zeta_{\pm}(a,b,-z)\nonumber\\&\pm&\frac{a}{2}z\int_{0}^{1}\frac{g(a,u)}{H^2(a,u)}\exp(-b/u)\zeta_{\pm}(a,b,u)\frac{du}{u+z}.
\end{eqnarray}
We have introduced in the right-hand side
\begin{equation}
\label{eqA14}
M_{\pm}(a,b):=q(a,b)\zeta_{\pm}(a,b,1/k(a)), 
\end{equation}
where
\begin{equation}
\label{eqA15}
q(a,b):=\frac{1}{2}\frac{R(a)}{H^2(a,1/k(a))}\exp(-k(a)b), 
\end{equation}
\begin{equation}
\label{eqA16}
R(a):=\frac{k(a)}{T'(a,1/k(a))}=\frac{1-k^{2}(a)}{k^{2}(a)+a-1},
\end{equation}
and the function $g$ in the integral term is
\begin{equation}
\label{eqA17}
g(a,u):=\frac{1}{T^+(a,u)T^-(a,u)}=\frac{1}{T^2(a,u)+(\pi a u/2)^2}\;\;(0\leq u<1).
\end{equation} 
Let $z$ go to infinity in both members of Eq. (\ref{eqA13}), first in the left complex half-plane, then in the right one. 
Writing $\zeta_{\pm}(a,b,-\infty)$ and $\zeta_{\pm}(a,b,+\infty)$ the limits obtained in this way, we have
\begin{equation}
\label{eqA18}
\zeta_{\pm}(a,b,-\infty)=(1\mp1)\zeta_{\pm}(a,b,+\infty).
\end{equation}
Consequently
\begin{equation}
\label{eqA19}
\zeta_-(a,b,+\infty)\zeta_+(a,b,+\infty)=1,
\end{equation}
due to (\ref{eqA12}).\\
The numerical calculation of the functions $\zeta_{\pm}$ is carried out starting from Eq. (\ref{eqA13}), which consists of 
two Fredholm integral equations on $[0,1]$. A difficulty arises because the coefficients $M_{\pm}(a,b)$ contain the unknown 
values of $\zeta_{\pm}$ at $1/k(a)$. It is overcome by putting $z=1/k(a)$ into Eq. (\ref{eqA13}) to calculate $M_{\pm}(a,b)$, 
which is next substituted into the right-hand side of (\ref{eqA13}). The expression of the functions $\zeta_{\pm}$ in 
the right half-plane becomes \cite{rutily-chevallier-bergeat2004}  
\begin{equation}
\label{eqA20}
\zeta_{\pm}(a,b,z) = \rho_{\pm}(a,b,z)\pm M_{\pm}(a,b)\sigma_{\pm}(a,b,z),\;(\Re(z)\geq 0),
\end{equation}
where $\rho_{\pm}$ and $\sigma_{\pm}$ are solution to the integral equations
\begin{equation}
\label{eqA21}
\rho_{\pm}(a,b,z) =  1\pm \frac{a}{2}z\int_{0}^{1}\frac{g(a,u)}{H^{2}(a,u)}\exp(-b/u)\rho_{\pm}(a,b,u)\frac{du}{u+z}\,, 
\end{equation}
\begin{equation}
\label{eqA22}
\sigma_{\pm}(a,b,z) = \frac{2k(a)z}{1+k(a)z}\pm\frac{a}{2}z\int_{0}^{1}\frac{g(a,u)}{H^{2}(a,u)}\exp(-b/u)\sigma_{\pm}(a,b,u)\frac{du}{u+z} \,.
\end{equation}
Once they have been solved on $[0,1]$, these equations define the functions $z \to \rho_{\pm}(a,b,z)$ and $z\to \sigma_{\pm}(a,b,z)$ 
everywhere in the right complex half-plane, in particular at $z = 1/k(a)$. Whence the coefficients $M_{\pm}(a,b)$ from 
Eqs. (\ref{eqA14}) and (\ref{eqA20}) are
\begin{equation}
\label{eqA23}
M_{\pm}(a,b) = \frac{q(a,b)\rho_{\pm}(a,b,1/k(a))}{1\mp q(a,b)\sigma_{\pm}(a,b,1/k(a))},
\end{equation}
$q(a,b)$ being defined by (\ref{eqA15})-(\ref{eqA16}).\\
Calculating the functions $\zeta_{\pm}$ thus requires the numerical solution of Eqs. (\ref{eqA21})-(\ref{eqA22}). In fact, 
it can be proved that the solution of Eqs. (\ref{eqA22}) comes down to that of Eqs. (\ref{eqA21}). This means that the 
functions $\sigma_{\pm}$ are analytically expressible in terms of the functions $\rho_{\pm}$. It follows that only Eqs. 
(\ref{eqA21}) have to be solved numerically in view of computing the functions $\zeta_{\pm}$. The details are not given here, 
since there is no inconvenience, at the level of accuracy, to solve both sets of equations numerically. This results from the 
regularity of the free terms and kernels of Eqs. (\ref{eqA21})-(\ref{eqA22}) on $[0, 1]$ and $[0,1]\times [0,1]$ respectively. 
Their solution $\rho_{\pm}$ and $\sigma_{\pm}$ are easily computable since they are regular and smooth on $[0,1]$, i.e., 
not only continuous, but differentiable everywhere in $[0,1]$, including at 0. We plan to publish ten-figure tables of these 
functions in the future.

\setcounter{equation}{0}
\setcounter{section}{2}
\setcounter{subsection}{0}

\section*{Appendix B. The functions $F_{\pm}=F_{\pm}(a,b,\tau)$}

These functions are defined by the relation
\begin{multline}
\label{eqB1}
F_{\pm}(a,b,\tau):=-\frac{1}{2}[\delta(\tau)\pm\delta(b-\tau)]-\frac{1}{2i\pi}\int_{c-i\infty}^{c+i\infty}H(a,1/z)\zeta_{\pm}(a,b,-1/z)\\
\times [\exp(\tau z)\pm\exp((b-\tau)z)]\frac{dz}{z} 
\end{multline}
where $-k(a)<c<0$. $\delta(\tau)$ and $\delta(b-\tau)$ are the Dirac distributions at 0 and $b$, respectively.

\subsection{Three properties of the functions $F_{\pm}$}
We first prove the following relations:
\begin{equation}
\label{eqB2}
F_{\pm}(a,b,b-\tau)=\pm F_{\pm}(a,b,\tau),
\end{equation} 
\begin{equation}
\label{eqB3}
F_{\pm}(a,b,0)=\frac{1}{\sqrt{1-a}}\zeta_{\pm}(a,b,+\infty)-1,
\end{equation} 
\begin{equation}
\label{eqB4}
(1-a)[1+F_{+}(a,b,0)][1+F_{-}(a,b,0)]=1.
\end{equation} 
Equation (\ref{eqB2}) is evident and Eq. (\ref{eqB4}) is a direct consequence of Eqs. (\ref{eqB3}) and (\ref{eqA19}). To 
derive (\ref{eqB3}), we note that
\begin{equation}
\label{eqB5}
F_{\pm}(a,b,0)=-\frac{1}{2}-\frac{1}{2i\pi}\int_{c-i\infty}^{c+i\infty}H(a,1/z)\zeta_{\pm}(a,b,-1/z)[1\pm\exp(bz)]\frac{dz}{z}.
\end{equation} 
The integral is transformed with the help of the residue theorem applied to the contour of Fig. 3, within which the integrand is analytic.
\begin{figure}[htb]
\centering
\includegraphics{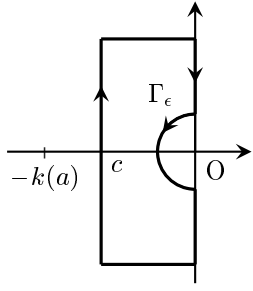}
\label{fig3}
\caption{Contour for the calculation of $F_\pm(a,b,0)$.}
\end{figure}
Since the integral along the half-circle $\Gamma_{\epsilon}$ tends to $(1/2)H(a,\infty)\zeta_{\pm}(a,b+\infty)(1\pm1)$ when 
$\epsilon \to 0$, we obtain
\begin{multline}
\label{eqB6}
F_{\pm}(a,b,0)=-\frac{1}{2}+\frac{1}{2}(1\pm1)H(a,\infty)\zeta_{\pm}(a,b,+\infty)\\
-\frac{1}{2 i \pi}\int_{-i\infty}^{+i\infty}H(a,1/z)(\zeta_{\pm})^+(a,b,-1/z)[1\pm\exp(bz)]\frac{dz}{z}.
\end{multline}
Put
\begin{align}
(\alpha)&=-\frac{1}{2i\pi}\int_{-i\infty}^{+i\infty}H(a,1/z)(\zeta_{\pm})^+(a,b,-1/z)\frac{dz}{z},\nonumber\\
(\beta)&=\mp\frac{1}{2i\pi}\int_{-i\infty}^{+i\infty}H(a,1/z)(\zeta_{\pm})^+(a,b,-1/z)\exp(bz)\frac{dz}{z},\nonumber
\end{align}
so that the integral term in (\ref{eqB6}) is $(\alpha)+(\beta)$. From Eq. (\ref{eq25}) with $z_0=1/z$, we have
\begin{align}
(\beta)&= -\frac{1}{2i\pi}\int_{-i\infty}^{+i\infty}H(a,-1/z)(\zeta_{\pm})^+(a,b,1/z)\frac{dz}{z}\nonumber\\
&\qquad\qquad+\frac{1}{2i\pi}\int_{-i\infty}^{+i\infty}H(a,-1/z)(\zeta_{\pm})^-(a,b,1/z)\frac{dz}{z},\nonumber\\
&=+\frac{1}{2i\pi}\int_{-i\infty}^{+i\infty}H(a,1/z)(\zeta_{\pm})^+(a,b,-1/z)\frac{dz}{z}\nonumber\\
&\qquad\qquad+\frac{1}{2i\pi}\int_{-i\infty}^{+i\infty}H(a,-1/z)(\zeta_{\pm})^-(a,b,1/z)\frac{dz}{z}.\nonumber
\end{align}
Therefore
\begin{equation}
(\alpha)+(\beta)=+\frac{1}{2i\pi}\int_{-i\infty}^{+i\infty}H(a,-1/z)(\zeta_{\pm})^-(a,b,1/z)\frac{dz}{z}.\nonumber
\end{equation}
Now integrate the function $z \to H(a,-1/z)\zeta_{\pm}(a,b,1/z)/z$ along the contour of Fig. 4. When $\epsilon \to 0$, the 
integral over the half-circle $\Gamma_{\epsilon}$ tends to $-(1/2)H(a,\infty)\zeta_{\pm}(a,b,-\infty)=-(1/2)H(a,\infty)(1\mp1)\zeta_{\pm}(a,b,+\infty)$ 
owing to Eq. (\ref{eqA18}), and the integral over $\Gamma_R$ tends to 1/2 when $R$ goes to infinity. Hence
\begin{equation}
(\alpha)+(\beta)=\frac{1}{2}(1\mp1)H(a,\infty)\zeta_{\pm}(a,b,+\infty)-\frac{1}{2},\nonumber
\end{equation} 
and Eq. (\ref{eqB6}) leads to (\ref{eqB3}) since $H(a,\infty)=1/\sqrt{1-a}$ from Eq. (\ref{eqA9}).\\
\begin{figure}[htb]
\centering
\includegraphics{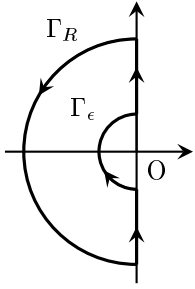}
\label{fig4}
\caption{Contour for the calculation of $(\alpha) + (\beta)$.}
\end{figure}

\subsection{Numerical calculation of $F_{\pm}(a,b,\tau)$}
It is carried out by replacing $H(a,1/z)$ by $1/[T(a,1/z)H(a,-1/z)]$ in the integral of (\ref{eqB1}), according 
to (\ref{eqA8}). This reveals the pole $z=-k(a)$ of the integrand in the left complex half-plane, together with the cut 
$]-\infty,-1]$ insuring that the function $z \to T(a,1/z)$ is single-valued in this half-plane. This function is 
then sectionally analytic in the complex plane cut along $]-\infty,-1[\,\cup\,]+1,+\infty[$, with limits $T^{\pm}(a,1/u)$ on 
the cut given by (\ref{eqA2}). Next, the residue theorem is applied to the contour of Fig. 5, which passes round the cut. 
\begin{figure}[htb]
\centering
\includegraphics{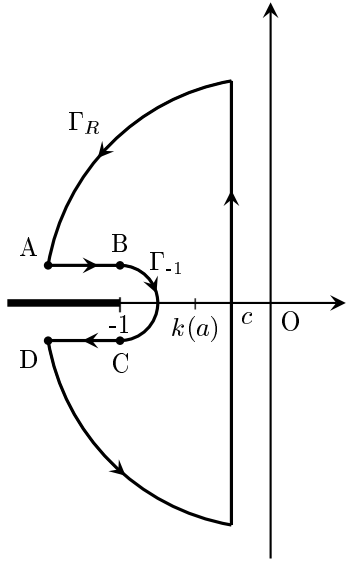}
\label{fig5}
\caption{Contour for the calculation of the functions $F_\pm$.}
\end{figure}
The unique pole within the contour is $z=-k(a)$, with residue 
\begin{equation}
[R(a)/H(a,1/k(a))]\zeta_{\pm}(a,b,1/k(a))[\exp(-k(a)\tau)\pm\exp(-k(a)(b-\tau))],\nonumber 
\end{equation}
since $R(a)$ is defined by (\ref{eqA16}). The integral on the half-circle $\Gamma_R$ tends to $[\delta(\tau)\pm\delta(b-\tau)]/2$
when $R \to \infty$, the integral on $\Gamma_{-1}$ tends to 0 with the radius, and the limit of the integral over AB+CD is
\begin{equation}
\frac{a}{2}\int_{1}^{+\infty}\frac{g(a,1/v)}{H(a,1/v)}\zeta_{\pm}(a,b,1/v)[\exp(-\tau v)\pm\exp(-(b-\tau)v)]\frac{dv}{v^2},\nonumber
\end{equation} 
where $g$ is defined by Eq. (\ref{eqA17}). Putting $u=1/v$ in the above integral, one obtains
\begin{multline}
\label{eqB7}
F_{\pm}(a,b,\tau)=\frac{R(a)}{H(a,1/k(a))}\zeta_{\pm}(a,b,1/k(a))\\\qquad\qquad\qquad\qquad\qquad\times[\exp(-k(a)\tau)\pm\exp(-k(a)(b-\tau))]\\
+\frac{a}{2}\int_{0}^{1}\frac{g(a,u)}{H(a,u)}\zeta_{\pm}(a,b,u)[\exp(-\tau/u)\pm\exp(-(b-\tau)/u)]du.
\end{multline}
For given functions $k(a)$, $H(a,u)$ and $\zeta_{\pm}(a,b,u)$, the numerical evaluation of the functions $F_{\pm}$ by means of 
this expression is easy, except when $a\to 1$ as regards the integrated term, and when $\tau \to 0^+$ or $b^-$ as regards the integral term. 
These difficulties are overcome in \cite{chevallier-rutily2004}.   

\paragraph{Acknowledgements}
We gratefully thank Profs. M. Ahues and A. Largillier (Universit\'{e} Jean Monnet, Saint-Etienne, France) for making several
useful comments on this work. We also thank Profs. J. Bergeat and G. Paturel (Observatoire de Lyon) for their careful reading 
of the manuscript.


\begin{thebibliography}{aa}
\bibitem{ahuesetal2002a} M. Ahues, F. D'Almeida, A. Largillier, O. Titaud, P. Vasconcelos, An $L^1$ refined projection
approximate solution of the radiation transfer equation in stellar atmospheres, Journal of Computational and Applied 
Mathematics {\bf 140}, 13-26 (2002).
\bibitem{ahuesetal2002b} M. Ahues, A. Largillier, O. Titaud, The roles of weak singularity and the grid uniformity in
relative error bounds, Numerical Functional Analysis and Optimization {\bf22}, 789-814 (2002).
\bibitem{rutily2002} B. Rutily, Multiple scattering theory and integral equations, in {\it Integral Methods in Science 
and Engineering}, C. Constanda, M. Ahues, and A. Largillier (eds.), Birkh\"{a}user, Boston (2004).
\bibitem{rutily-bergeat1994} B. Rutily, J. Bergeat, The solution of the Schwarzschild-Milne integral equation in an homogeneous
isotropically scattering plane-parallel medium, Journal of Quantitative Spectroscopy and Radiative Transfer {\bf51}, 823-847 
(1994).
\bibitem{chevallier-rutily2004} L. Chevallier, B. Rutily, Exact solution of the standard transfer problem in a stellar 
atmosphere. Journal of Quantitative Spectroscopy and Radiative Transfer {\bf91}, 373-391 (2005).
\bibitem{muskhelishvili1992} N.I. Muskhelishvili, Singular integral equations, OGIZ, Moscow-Leningrad (1946) [Noordhoff, 
Groningen, Holland (1953); Dover, New York (1992)].
\bibitem{busbridge1955} I.W. Busbridge, On the $X$- and $Y$-functions of S. Chandrasekhar, Astrophysical Journal {\bf122}, 
327-348 (1955).
\bibitem{mullikin1964} T.W. Mullikin, Chandrasekhar's $X$ and $Y$ equations, Transactions of the American Mathematical Society {\bf113}, 316-332 (1964).
\bibitem{carlstedt-mullikin1966} J.L. Carlstedt, T.W. Mullikin, Chandrasekhar $X$- and $Y$-functions, Astrophysical Journal Suppl.
{\bf12}, 449-585 (1966).
\bibitem{busbridge1960} I.W. Busbridge, The mathematics of radiative transfer, Cambridge University Press, London (1960).
\bibitem{rutily-bergeat2002} B. Rutily, J. Bergeat, Radiative transfer in plane-parallel media and Cauchy integral equations. I. 
The $N$-function, Transport Theory and Statistical Physics {\bf31}, 659-671 (2002).
\bibitem{rutily-bergeat-chevallier2004} B. Rutily, J. Bergeat, L. Chevallier, Radiative transfer in plane-parallel media and Cauchy 
integral equations. II. The $H$-function. Submitted for publication. 
\bibitem{rutily-chevallier-bergeat2004} B. Rutily, L. Chevallier, J. Bergeat, Radiative transfer in plane-parallel media and 
Cauchy integral equations. III. The finite case. Submitted for publication.


\end{thebibliography}
\end{document}